\documentclass[epj]{svjour}

\usepackage{graphicx}
\usepackage{dcolumn}
\usepackage{bm}
\usepackage{epsfig}

\newcommand{\rev}[1]{{#1}} 
\newcommand{\revrev}[1]{{#1}} 
\begin{document}

\title{Collapsing granular suspensions}

\author{D. Kadau\inst{1} \and J. S. Andrade Jr.\inst{1,2} \and
H. J. Herrmann\inst{1,2}}

\institute{
\inst{1} IfB, HIF E12, ETH H\"onggerberg,
8093 Z\"urich, Switzerland\\ 
\inst{2} Departamento de F\'{\i}sica,
Universidade Federal do Cear\'a, 60451-970 Fortaleza, Cear\'a, Brazil}

\date{Received: date / Revised version: date}

\abstract{
A 2D contact dynamics model is proposed as a microscopic
description { of a collapsing suspension/soil} to capture the essential
physical processes underlying the 
dynamics of generation and collapse of the system. Our physical model is
{ compared with} real data obtained from {\it in situ} measurements performed
with a natural 
{ collapsing/suspension soil.} 
We show that the shear strength behavior of our
collapsing suspension/soil model is { very similar to} the behavior of this
{ collapsing suspension soil,}
for both the unperturbed and the perturbed phases of the
material.
}


\maketitle

\section{Introduction}

{
In nature, fragile or metastable granular structures, collapsing under an
applied load exist in a large variety. 
 For instance, these collapsing suspensions/soils are well known from soil
mechanics as an unsaturated (by water) grain structure which collapses when
adding water (thus becoming saturated) or/and when loaded.
 They are also known to be loose structures and can be water deposited or aeolian
\cite{Mitchell05,Reznik05,Ayadat07}.
Similar structures are also responsible for collapsible behavior of colloidal
gels \cite{Manley05} or even snow \cite{Heierli05}.
}

In the present { work} 
we
develop and numerically solve a simple physical model for 
a collapsing suspension/soil representation, that is {capable
of incorporating} at the microscopic level the essential structural and
dynamical features of the material when subjected to different types
of perturbations. Within this context we focus on two aspects of the
problem, namely the penetration of {an intruder}, \rev{ leading to a partial
collapse of the system,} and the shear resistance
property of the material. { The model results of our approach are then
  compared to real and {\it in situ} data 
  of  a collapsing suspension/soil.}

\section{Simulation method}
The discrete element method constitutes a general class of modeling
techniques to simulate the microscopic behavior (i.e., at the particle
scale) of granular/soil materials. Here a variant of the contact
dynamics method, originally developed to model compact and dry systems
with lasting contacts \cite{Jean92}, is used to describe a 2D cohesive
suspension/soil material. This technique is conceptually based on the
exact implementation of non-smooth contact laws, which means that the
steric volume exclusion for perfectly rigid particles and the Coulomb
friction law are strictly implemented
\cite{Unger2003,Brendel2005}.

The absence of cohesion between particles can only be justified in dry systems
on scales where the cohesive force is weak compared to the gravitational force
on the particle, i.e. for dry sand and coarser materials, which can lead to
densities close to that of random dense packings. For collapsing
suspensions/soils, an attractive force must play an important role in the
stabilization of large voids \cite{Kadau03,Taboada2006}, leading to highly
porous systems as e.g. in fine cohesive powders. In the nanometer range of
particle sizes, the cohesive force becomes the dominant force, so that
particles stick together upon first contact. Here the bonding between two
particles is considered in terms of a cohesion model with a constant
attractive force $F_c$ acting within a finite range $d_c$, so that for the
opening of a contact a finite energy barrier $F_cd_c$ must be overcome.
{The utilization of this simplified cohesion model is justified in
  the case of short range interactions as we are not interested in the
  detailed behavior of an individual contact, but on the collective behavior
  of the system, where the opening of a contact is mostly determined by the
  force and energy barrier. More precisely, using a more complex model would
  not lead to a drastic change in the macroscopic behavior when keeping the
  same force and energy barrier. In most force-driven simulation as in this
  paper the cohesion force dominates the behavior and the energy barrier plays
  a minor role but its existence is of crucial importance to avoid simulation
  artefacts \cite{Kadau03,Brendel2005}.}  As we show later, this cohesion
force can be mediated, for example, by bacteria present in the system
{ which are much smaller than the grains, i.e. we have an effective
  short range interaction. We use $d_c=10^{-4}$ (in units of particle radii)
  for all the following simulations.  In addition, we implement friction and
  rolling friction between two particles in contact, so that large pores can
  be stable in the system. For comparison, we adopt the same friction
  coefficient $\mu=0.3$ and coefficient for rolling friction $\mu_r=0.1$ (in
  units of particle radii) in all simulations, taking into account that
  rolling of particle contacts is easier than sliding
  \cite{Kadau03,Brendel2005}. }

In the case of collapsing suspensions/soils one also has to take into
account the time necessary for bonds to appear, i.e.\ during
relatively fast processes new bonds will not be formed, whereas for
longterm processes bonds are allowed to form at a particle
contact. Finally, gravity also cannot be neglected in the model since
the particle diameter is usually well above the micron-size. For
simplicity, however, the surrounding pore water is not explicitly
considered but only taken into account as a buoyant medium, reducing
the effective gravity acting onto the grains.  In this way, we
disregard fluid motion.{ This hypothesis should remain valid for the
  settling 
grains in the system  as long as
the velocities reached by each particle  are
sufficiently large (i.e., inertial effects are relevant), and the
total virtual buoyancy induced by all particles released during the
penetration process is small \cite{Noh00}. The issue of disregarding the fluid
motion will be discussed in more detail when presenting the results in the
following section.}

\section{Results}
\rev{
As mentioned earlier the major feature of the metastable granular structures,
investigated in this paper is that they collapse, e.g.\ when an external load
is applied. This will be illustrated and investigated in section \ref{sub:intruder} by penetration
of an intruder leading to partial collapse of the structures.
Understanding the penetration behavior will also be important for the
experiments presented in section \ref{sub:exp}, where the measurement
device has to be pushed inside the soil before the measurements.
Before and after the collapse the shear strength of the material changes
drastically as shown in the simulations  (sec.\ \ref{sub:shear}) and in
experiments (sec.\ \ref{sub:exp}).  

A useful number for characterization of the structures after the (partial)
collapse where many bonds break is the fraction of still active bonds
(i.e.\ fully cohesive) with respect to the total number of contacts. This
number will be  
used in the following for interpretation of the results.}


\subsection{Penetration of an intruder}\label{sub:intruder}
\begin{figure}
a \includegraphics[width=2.5cm]{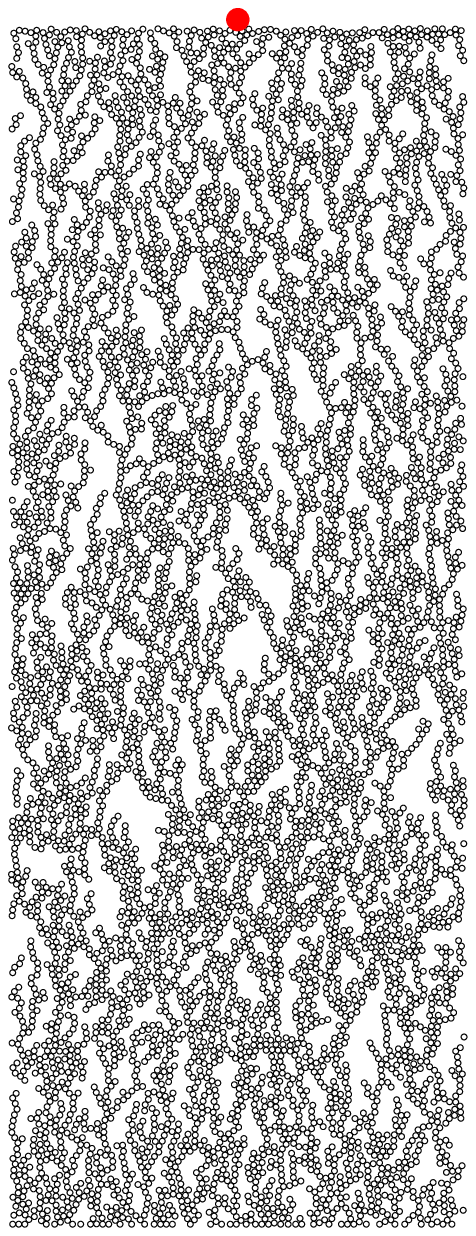} 
b \includegraphics[width=2.5cm]{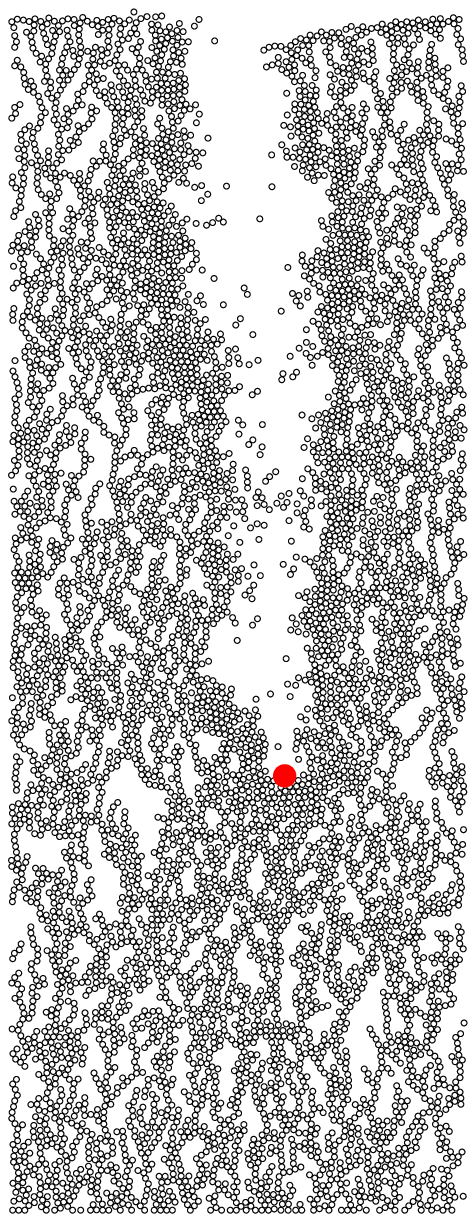} 
c \includegraphics[width=2.5cm]{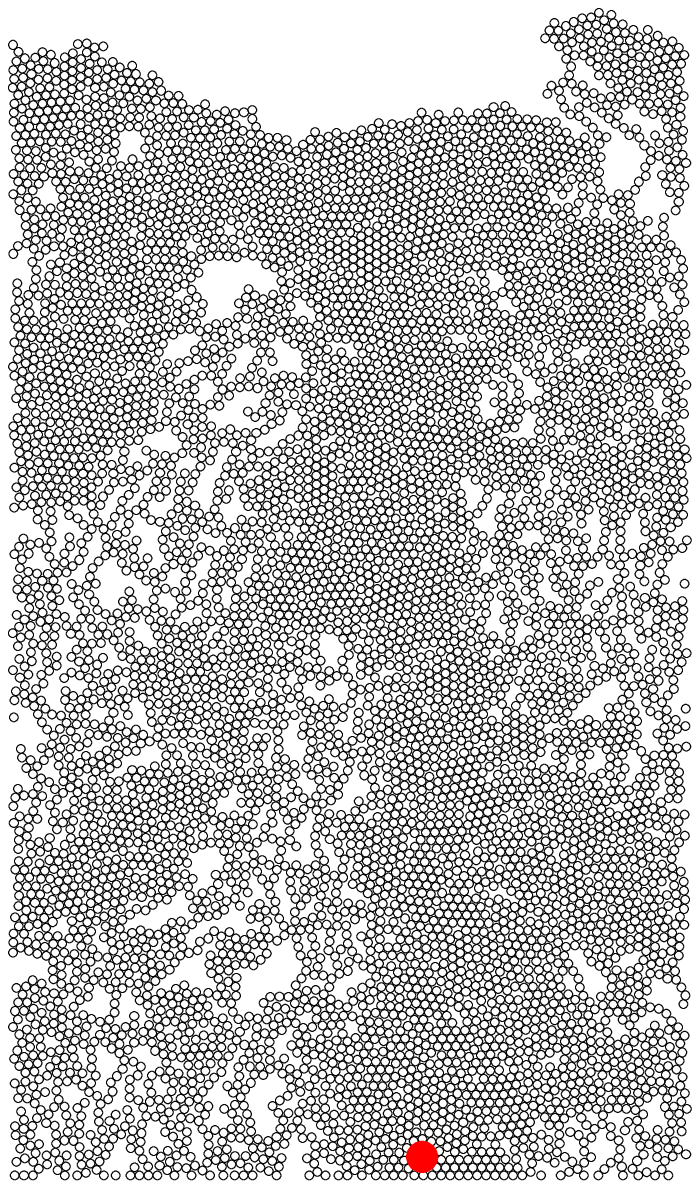} 
\caption{\label{fig:snaps}
Snapshots from simulations showing a typical realization of the
penetration process of an intruder into a very loose cohesive packing.
The (unperturbed) collapsing suspension/soil shown in (a) is modeled
as a tenuous granular network of cohesive disks assembled by a gravity
driven process of ballistic deposition and contact dynamics
\cite{Jean92,Kadau03}. {The system size is $51\times 199$ in units of grain
diameters, corresponding to 1.7cm$\times$6.7cm when using the typical diameter
 $d=344\mu$m of the grains in the experiments.} As shown in (b), the movement
of the 
intruder is responsible for the partial destruction of the granular
structure along its trajectory. At the end (c), the intruder rests
under a compact mass of (perturbed) material.}
\end{figure}
%
{We carry out simulations of a large disk (intruder) of low density
  (half the grain density) pushed with constant force into the granular
  structure and subsequently removed again from a fragile cohesive granular
  structure.}  Only if the force exceeds a certain threshold it penetrates the
medium. Additionally, we assume that the time scale for the packing generation
is large enough to create bonds between the grains, whereas the movement of
the intruder is on time scales much smaller than the bonding time. The
sequence of snapshots shown in Fig.~\ref{fig:snaps} corresponds to different
stages of the penetration process into a typical realization of a very loose
packing. The initial (unperturbed) network structure (Fig.~\ref{fig:snaps}a)
is a result of the aggregation of cohesive disks by means of a gravity driven
process which includes ballistic deposition with contact dynamics
\cite{Jean92,Kadau03}.  {At this point we are mostly interested in
  the penetration process into a fragile structure independent on its process
  of generation. The specific configurations used here are generated to
  resemble as closely as possible the experiments described later (sec.\
  \ref{sub:exp}). The details
  of this generation process will be described later (sec.\
  \ref{sub:shear}). The initial configurations used here 
  (Fig.~\ref{fig:snaps}) are created with a value $F_c/F_g=10^4$ for the
  cohesion force $F_c$ normalized by the gravitational force $F_g$ on each
  grain.  Deliberately we chose this high value to obtain a rather loose
  structure \rev{ with coordination number slightly above 2 ($2.06$),} since lower values of $F_c/F_g$ lead to denser structures (cf.\
  sec.\ \ref{sub:shear}) with smaller pores, i.e.\ a
  more homogeneous microstructure as found \rev{for non-cohesive materials and
  therefore relatively dense materials}  in laboratory experiments
  \cite{Lohse04,Royer05,Caballero07}. 

  Note, that the microstructure of two
  dimensional systems gives, at first glance, a less homogeneous impression
  than the  corresponding three dimensional systems as pores are much easier
  to identify visually. This has been illustrated, e.g.\  for the compaction
  of cohesive powders when comparing two dimensional simulations
  \cite{Kadau03} to three dimensional simulations   \cite{Bartels05}.
 
 }
Once
the intruder breaks through, the highly porous material collapses
under the action of gravity. We observe the creation of a channel
(Fig.~\ref{fig:snaps}b) which finally collapses over the descending
intruder. At the end, the partial destruction of the network due to
the penetration process generates a large amount of compact
(perturbed) material over the intruder. As shown in
Fig.~\ref{fig:snaps}c, the larger disk is buried under the debris of
particles. 
\rev{In this case of a relatively ``strong penetration'' (cf.\ following
  paragraph) the fraction of active bonds for the total packing is about 30\%.
   This means that about 43\% of the cohesive contacts in the initial
   configuration remain intact as during the collapse new contacts are formed,
   leading to an increase in the average coordination number to about
   $2.82$. These numbers, averaged over the whole systems, can be seen as a
   measure for the extent of still intact areas and collapsed areas. In the 
   fully intact areas  still all bonds are active and
   the coordination number is unchanged, slightly above $2$. On the contrary,
   in the fully 
   collapsed area, i.e.\ the channel created by the intruder, only about 2\% 
   of the bonds are still active, and the coordination number is about
   $3.2$. In this respect the total numbers can also be viewed as an indicator
for the collapsed fraction of the structure. 

 The {results} presented above } clearly suggests that objects lighter than
water can be effectively swallowed in a 
{ collapsing suspension/soil.} 
Furthermore, if we assume that cohesive bonds can be
restored by some particular physico-chemical or biological mechanism,
the force needed to remove the intruder disk could be significantly
higher than the originally penetration force.
\begin{figure}
\includegraphics[width=0.95\columnwidth]{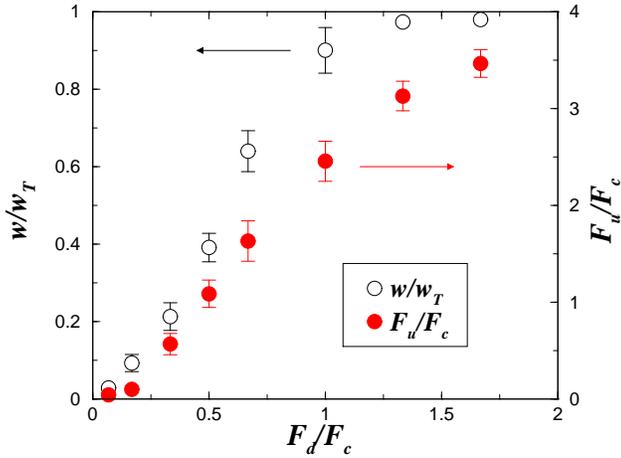}
\caption{\label{fig:weightforce} 
Dependence of the normalized weight of 
{ material} above the intruder
after penetration $w/w_{T}$ (open circles, left ordinate) on the
applied pushing force $F_{d}/F_{c}$ (normalized by cohesive force). In
the same plot, we also show how the normalized force necessary to pull
out the intruder $F_{u}/F_{c}$ (filled circles, right ordinate) varies
with $F_{d}/F_{c}$.}
\end{figure}
In order to provide a more quantitative confirmation for this
interesting behavior, several simulations of the intruder
penetration/removal process have been performed for $10$ different
realizations of granular networks generated with the same microscopic
cohesion and friction parameters. In Fig.~\ref{fig:weightforce} we
show the dependence of the weight fraction $w/w_{T}$ on the ratio
$F_{d}/F_{c}$, where $w$ is the weight of material above the intruder
after the penetration process, $w_{T}$ is the total weight of the
sample, $F_{d}$ is the force applied to push the disk down and $F_{c}$
is the characteristic value used for the cohesion force. 
 As depicted,
the force to pull up the intruder disk can indeed be as much as three
times higher than the force to push it down. 
\rev{ Note, that the
  fraction of active bonds in the final packing strongly depends on the
  ``penetration strength''. For low values of $F_d/F_c$, where the intruder is
  not pushed into the structure very deeply, most of the bonds
  are still  active. For $F_d/F_c\sim 0.33$, e.g.\  93\% of the bonds are
  active with respect to the final structure, corresponding to  97\% of the
  original bonds of the initial structure, i.e.\ very few
  bonds are broken. Also, for $F_d/F_c\sim 1$, where the intruder almost reaches
  the bottom, still 68\% of the bonds are active (78\% of the original bonds).
  As the intruder can only go down until reaching the
 bottom, increasing the ``penetration strength'' further only leads to the
 breaking of more cohesive bonds. This indicates a stronger collapse of the
 structures which explains why the the force to pull up the intruder still
 increases for high values of  $F_d/F_c$.  } 

{

  The results presented above illustrate the typical behavior of such a
  fragile structure for a given initial density (here the initial volume
  fraction is about $0.43$). Changing the volume fraction (density), the
  curves are shifted to the right for higher densities or to the left for
  lower densities, still showing the same qualitative behavior.} 
 {We expect three dimensional simulations to display a
  qualitative similar penetration behavior, but the intruder experiencing less
  resistance of the material due to the additional degree of freedom. This
  effect has been found when comparing the pore stabilization mechanisms of
  two dimensional and three dimensional simulations \cite{Kadau03,Bartels05}.}
{Disregarding the interstitial fluid motion keeps the model as simple
  as possible and nevertheless able to reproduce the main experimental
  observations. Of course the details of the collapse of the material may be
  influenced by the flow field of the surrounding fluid
\rev{  (e.g.\ \cite{Caballero07,Royer05}).}  
  We tested the influence of the fluid by introducing a viscous drag
on the grains. For this, the drag coefficient of water and a typical
grain size of $d=344~\mu$m have been used. The simulation results
showed no significant difference. }

\rev{ The results of this section are also very important for the experiments
  presented in section \ref{sub:exp} using a vane rheometer pushed into the
  investigated soil.  When pushed in smoothly the thin vane
blades will not damage too much the structure
for the material to be considered unperturbed.
 This is one reason for choosing a
  relatively small intruder, being only three times larger than the grains,
  and being in the range or even smaller than the typical pore size of the
  structure. Additionally, this saves a lot of computation time as for larger
  intruders also the system sizes have to be larger. Preliminary studies with
  larger intruders show similar qualitative behavior and will be the subject of
  future studies.  }

\subsection{Shear Resistance} \label{sub:shear}

{

  The unperturbed material} is modeled by a
ballistic deposition of particles driven by gravity, { as an example of a
  process that 
generates very fragile structures.}  During this
relatively slow process, bonds are allowed to form when particles
stick to each other. { As previously discussed the origin of the
attractive force can be explained  in terms of cohesive
bonds  mediated by the bacteria living in the
suspension \cite{Kadau08}. }
{For the bonds, we use the same model as described before, i.e.\ the
  bond strength is $F_c$, which is the dominant parameter determining the
  density of the system.}
The particles are sequentially deposited with a fixed time
interval between each one to allow for relaxation. 
To get a very loose and fragile structure, an
extremely large time between two depositions should be used. 
 The procedure adopted
here is to gradually increase the time interval. We only stop when the
packing generated with the highest time interval has, within a given
tolerance, the same particle density of the previous one.  We show in
Fig.~\ref{fig:simpertubation}a that this procedure results in a highly
porous, and therefore tenuously connected, network of grains.
{In
  practice, for the simulations presented here, depositing one particle
  each time step turned out to be already adiabatic, 
   i.e.\ the density did not change any more (less than
  0.1\%).
{ Note, that the generation process of the fragile structure used in our
  simulations somehow resembles 
  the processes of formation 
 of  natural 
 collapsing suspensions/soils investigated in  {\it in-situ} experiments
\cite{Kadau08}. This will later be important to be able to compare the experiments
with our model (sec.\  \ref{sub:exp}).}

\begin{figure}
\begin{center}
\hspace{ -1em} a \includegraphics[width=3.5cm]{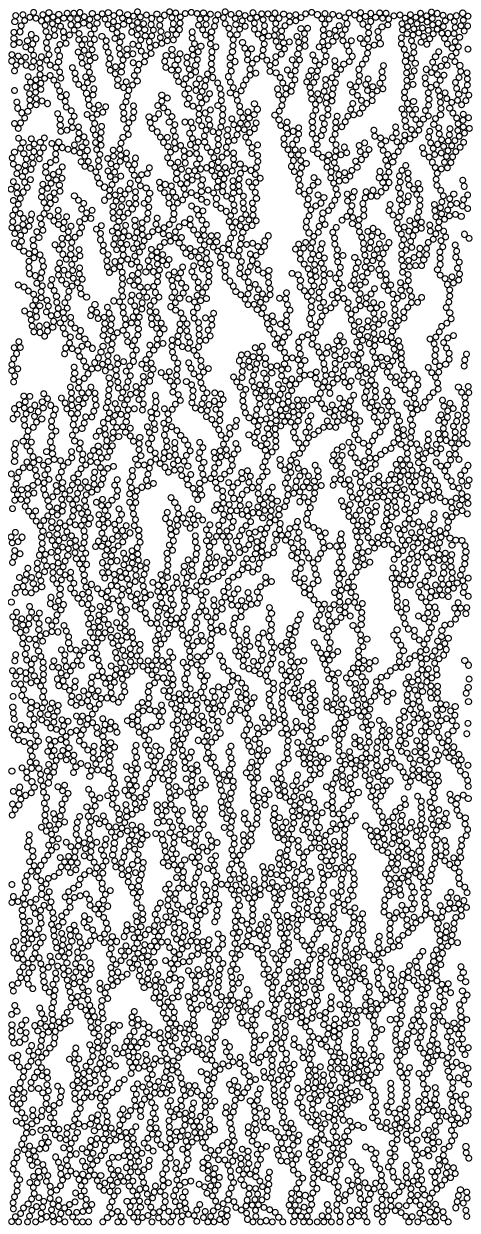}
b \includegraphics[width=3.5cm]{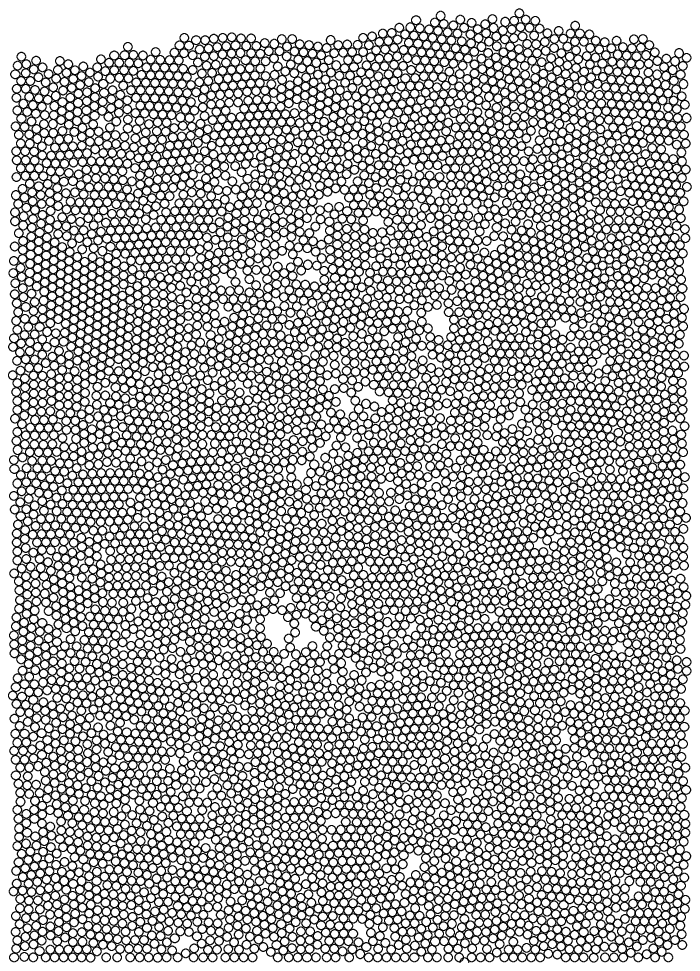} 
\end{center}
\caption{\label{fig:simpertubation} Typical realizations of the 
granular system prepared for computational simulations of shear
strength {(periodic boundaries in horizontal direction).} In (a) we show the (unperturbed) granular network of
cohesive disks {($F_c/F_g=10^4$)} generated by ballistic deposition and contact dynamics,
and in (b) the (perturbed) compact structure obtained after the
collapse of the system due to a piston-like pushing force applied at
the top {The system size is $51\times 199$ in units of grain
diameters, corresponding to 1.7cm$\times$6.7cm when using the typical
diameter $d=344\mu$m of the grains in the experiments, b: $51\times 106$}.}
\end{figure}
%

{ With the purpose to simulate the shear strength 
after the collapse,
} we
implement in our model a 
piston-like pushing force acting on the particles at the top. In this
situation, the collapse of the granular network that starts from the
upper region of the system is analogous to a gradual increase in the
effective piston weight, leading to an acceleration of the compaction
mechanism. During this process
bonds will break without reforming again up to the point in which most
of the bonds are broken (see Fig.~\ref{fig:simpertubation}b).
{ This behavior will be valid for all materials where the time scale of
  bond formation is larger than that of the collapsing and measurement, or
  where the bonds are not able to reform again at all,  as   e.g.\  in the {\it in situ}
experiments { (sec.\ \ref{sub:exp})}.}
\rev{ After the collapse only 8.5\% of the bonds are active (i.e.\ cohesive
  contacts) corresponding to about 13\% of the active bonds of the initial
  structure,   i.e.\    most of the bonds are broken. 
  As the number of contacts increases during the collapse the coordination
  number increases from slightly above 2 ($2.05$) to about $3.17$, whereas
  the volume fraction increases from about $0.4$ to about $0.77$. 
   The value for the coordination number is almost the same as within the
  channel  after the
  compaction by an intruder (cf.\ sec.\ \ref{sub:intruder}: coordination number
  about $3.2$), whereas the fraction of active bonds, although very low, is
  still higher  than in the "intruder" case.  This can be understood  because
  the intruder destroys the bonds very efficiently on its way down. Then, the
  material settles again without almost no active bonds in the ``channel''. In
  the case of compacting the whole system ``externally'' as described above
  not all 
  bonds have to be broken to compact the system.  
  For the case of very large intruders one expects both situtions: One part of
  the system 
  below the intruder is compacted similar as the whole system. Another
  part of the system collapses over the intruder within its ``channel''. 
}

 We
calculate the shear strength of the unperturbed and perturbed
structure as follows. At a given depth, we apply a constant force in
the horizontal direction to a randomly chosen particle and observe how
far it can move in this direction.  Our assumption here is based on 
the fact that in real experiments the rheometer creates a thin shear
layer at the upper and lower edges of the vane. This layer is
represented in our model by these sample particles subjected to a
constant force. By changing the force using nested intervals, one can
calculate within a given numerical tolerance the minimum force
necessary to move this particle at a distance that is sufficiently
large to sample the disordered porous geometry (e.g., approximately
$20$ particle diameters away). This procedure is then repeated for
different particles at the same depth to produce an average shear
strength value. { During this procedure, measuring the threshold
in a static system when motion sets in, it is justified to disregard the fluid motion.}

\begin{figure}
\includegraphics[width=0.95\columnwidth]{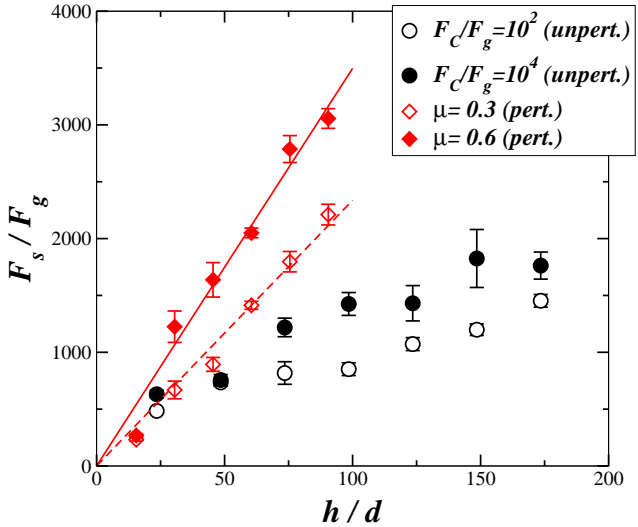}
\caption{\label{fig:simshear}  
{Dependence of the simulated shear strength $F_s$ (normalized by the
gravitional force $F_g$ of each grain) on the depth for
unperturbed and perturbed systems (depth $h$ in units of grain diameters
$d$).} { Compared to the perturbed system
(diamonds), the unperturbed grain structure shows only a weaker
dependence on the depth. { This is similar to the
behavior of a real collapsing suspension/soil (see Fig.~\ref{fig:visco}).} For the unperturbed system,  an increase
by a factor of $100$ in the cohesion force does not change
substantially the behavior of the shear strength. For the perturbed system we
also  show two different
values of the friction coefficient  $\mu$ ($0.3$ and $0.6$). A higher friction
coefficient leads to a stronger increasing shear strength as
  indicated by the slopes ($a\approx 23.4$ for $\mu=0.3$ and $a\approx 35$ for $\mu=0.6$) of the linear fit as in Eq.(\ref{eq1}).} }
\end{figure}
The results in Fig.~\ref{fig:simshear} show the variation with depth
of the shear strength averaged over $10$ different particles for both
unperturbed and perturbed systems. The cohesion forces are
dominant in the unperturbed state while the contribution of friction
forces governs the dynamical behavior of the perturbed material.  
{ The model results
obtained with the unperturbed system show only a minor dependence of
the shear strength on the depth compared to the perturbed
structure.  Furthermore, for the unperturbed system the shear strength does not
extrapolate to zero in the limit of low depths.}
 Additionally, we observe that a
significant increase in the cohesion force intensity (e.g., by a
factor of one hundred) does not lead to a substantially different
behavior of the shear strength. This is a consequence of two distinct
effects, namely that ({\it i}) the particle density decreases with the
cohesion force, and that ({\it ii}) a larger shear strength is
necessary to move a particle in a more compact medium (i.e., a high
density packing \rev{with volume fraction of about 0.69 and coordination
  number  about 3}). As also shown in Fig.~\ref{fig:simshear}, the shear
strength increases linearly with depth for the case of the perturbed
material.  Expectedly, as we observe in the same
plot, an increase in the friction coefficient generally increases the
average shear strength.
\revrev{
Using the slopes of Fig.~\ref{fig:simshear} one can for the perturbed case
calculate the effective friction coefficient of the material by simply
deviding the shear strength at a certain depth by the weight of the particles
above that depth. Here we use the full depth of the whole structure to achieve
highest accuracy. For our simulation this results in an effective friction
coefficient of $0.32$ ($\mu=0.3$) and $0.48$ ($\mu=0.6$). The relatively high
values and in particular the large difference can not solely be explained by
the presence of friction. It can better be understood by the fact that there are still active bonds present in the structure, enhancing the strength of the material.}
}



\subsection{Comparison to Experiments} \label{sub:exp}

{ In order to test our physical model
  we use experimental data from investigations of a specific type of a natural
collapsing suspension/soil  \cite{Kadau08}. This comparison between the real data and our contact dynamics model
can only be made qualitatively due to obvious difficulties in
obtaining reasonably precise experimental measurement of any {\it in
situ} microscopic parameters. 

 The experimental data come from a
natural reserve called Len\c{c}ois Maranhenses located in the
North-East of Brazil \cite{Kadau08}.}
 We found that at the shore of
many drying lakes in this place, it is common to find a special type
of { a collapsing suspension/soil} consisting of an impermeable crust lying
above a 
metastable suspension of grains \cite{Danin78}. 
 { The slow drying process of the lakes, including continuous deposition of
   grains transported by the wind from the adjacant dunes into the lakes,  is
   consistent with the slow limit of depositing grains used in the simulations
   (sec.\ \ref{sub:shear}) leading to very fragile structures.}
 Provided one
does not exert on the surface a pressure higher than $p_c =
10-20$~kPa, it is possible to step on it and the surface will
elastically deform in a very similar way to what happens when one
walks on a waterbed. These deformations visibly extend over a couple
of meters. 
If the pressure $p_c$ is exceeded, the surface cracks in a
brittle way producing a network of tensile cracks \cite{Kadau08}. 
After that we observe the separation of the excess pore-water from 
a repacked and wet sand soil \cite{Parker66} with pronounced shear 
thinning behavior \cite{Khaldoun05}. Because the collapse of this 
{ suspension/soil} is irreversible, we had to study its rheology
  and strength  
{\it in situ}. { This irreversible collapse leading to a compacted
  material with reduced pore volume, could be qualitatively reproduced in our
  simulations, as presented before \rev{ (cf.\ sec.\ \ref{sub:intruder} and  \ref{sub:shear}).}}

\begin{figure}
\vspace{2ex}
\includegraphics[width=0.95\columnwidth]{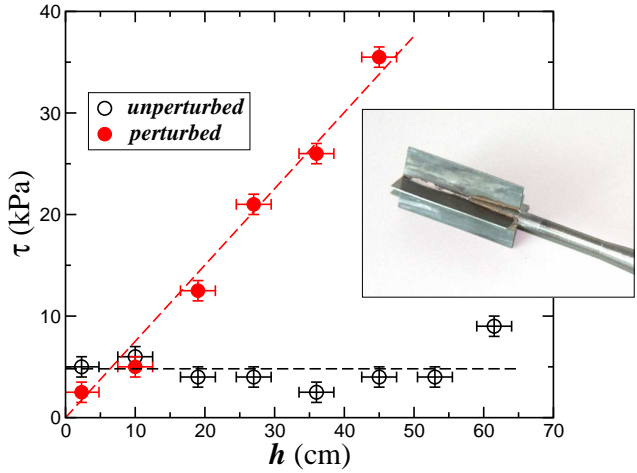}  
\vspace{2ex}
\caption{\label{fig:visco}
Experimental measurements of the shear strength  as a function of depth
before (open circles) and after (filled circles) the collapse of the
{ suspension/soil.} \rev{ For the measurements, we used a vane rheometer
  Geonor H-60 shown in the inset. } {The horizontal error bars indicate the read-off accuracy
  of the vane rheometer, the horizontal error bars are given by the height of
  the vane.} The least squares fit to data of a linear function
$\tau=ah$ gives $a=0.75 \pm 0.03$~kPa/cm after the collapse. The shear
strength of the unperturbed 
{ suspension/soil} (before the collapse) follows an
approximately constant behavior $\tau \approx 5$~kPa until reaching
the bottom of the system, which was at about 60~cm below the surface
for this case.}
\end{figure}

By placing light plates on the surface we could walk on the 
{ suspension/soil} without
visually modifying it and make various measurements before and after the
collapse. The most striking result concerns the shear strength $\tau$ measured
with a vane rheometer.
\rev{ The vane consists of 4 thin blades (see inset in fig.\ \ref{fig:visco}).
  During the measurement the vane rotates cutting out a cylindrical area
  of the material. 
 In this respect the results of  section \ref{sub:intruder} are very
 important, as the rheometer has to be pushed into the
 investigated soil, before measuring. The thin vane blades 
 ensure that when pushed in smoothly enough 
 most of the structure in the cylinder rotated is not
   not destroyed by the vane. Thus, we were really able to measure the properties of the unperturbed material.}

 { The vane rheometer measures both the
  threshold stress and the steady state stress. Here we measure the maximal
  torque necessary to rotate the vane and, knowing its geometry, it is
  possible to calculate the threshold stress below which the vane does not
  move \cite{Clayton95}.  As shown in Fig.~\ref{fig:visco}, before} destroying
the \rev{structure}, $\tau$ is essentially constant up to the bottom of the basin and
then it rapidly increases.
{ In order to reproduce this behavior with our model,
 simulations with much larger system sizes would be necessary.}
 After the system collapsed and the water came out,
$\tau$ linearly increases with depth $h$:
\begin{equation}
\tau(h) = a h~,
\label{eq1}
\end{equation}
with $a=0.75 \pm 0.03$~kPa/cm. We conclude from our measurements that
the { soil} 
essentially consists of a suspension with depth
independent \revrev{shear strength.} 
 After the collapse, it becomes a soil
dominated by the Mohr-Coulomb friction criterion for its shear
strength.
{ Such a behavior is again consistent with our model simulations of a
  collapsing soil/suspension. }
\rev{Summarizing, we found a cross-over from a yield stress material with a
  threshold value  independent on depth to a Coulomb material after the
  collapse. At least, in the experiments the depth-independent threshold value
  can be  clearly seen (fig.\ \ref{fig:visco}). } 

The analysis of this collapsing soil shows grains with a typical
size of $d=344~\mu$m and standard deviation $\sigma_d=120~\mu$m as
well as the presence of a huge amount of {\it cyanobacteria} and {\it
diatomacea} of various types \cite{Kadau08}. During the drying of the
lakes, these organisms are responsible for the formation of the
elastic and impermeable crust which prevents further water from
evaporating out of the metastable suspension of grains. As reported in
previous studies \cite{Danin78}, we confirmed that the cohesion force
between grains inside the suspension is mediated by the {\it
cyanobacteria} present in the system.


\section{Conclusion}


Our model is capable to simulate the most important
features of a real collapsing suspension/soil. { Most features are
  important for many  types of collapsing suspension/soil including
  wet and dry quicksand, 
 and the model
  may also be applied for the collapse of cohesive
  powders or snow.
}  Shed by bacteria in a
highly unstable granular skeleton, this 
 { suspension/soil} can
catastrophically collapse. During this rapid segregation, objects
lighter than water, once pushed deep down through the collapsing
suspension, can be irreversibly trap\-ped under the debris of
disassembled grains and clusters of grains. { According to
our simulation results this behavior appears  to be a general feature of a
collapsing suspension/soil. \rev{ In our case, the fraction of still active
  bonds with respect to the total number of contacts, is a very useful number
  for describing the collapsed structures microscopically. } 
Moreover, our model 
results  show similar behavior for the shear strength  of the
real material} in both unperturbed and perturbed phases, \rev{ showing a
  cross-over between a yield stress material to a Coulomb material
  after the collapse.}  As a challenge 
for future work, we intend to investigate the effect of size and shape
of the intruder on the ``drag'' force exerted by the granular network
\cite{Albert99} in the presence of cohesive forces.

\thanks{
This research was supported by a grant from the G.I.F., the German-
Israel Foundation for Scientific Research and Development. Furthermore, we thank CNPq, CAPES, FUNCAP, FINEP, the Volkswagenstiftung and the
Max Planck Prize for financial support.
}

\end{document}